\documentclass[preprint]{revtex4-1}
\usepackage{graphicx}
\usepackage{amsmath}
\usepackage[caption = false]{subfig}
\usepackage{siunitx}
\usepackage{xcolor}

\usepackage{graphicx}

\usepackage{amssymb}

\begin{document}

\title{{\it{Ab-Initio}} Surface Hopping and Multiphoton Ionisation Study of the Photodissociation Dynamics of CS$_2$}

\author{Darren Bellshaw,\textit{$^{a}$} Daniel A. Horke,\textit{$^{b,c}$} Adam D. Smith,\textit{$^{d}$} Hannah M. Watts,\textit{$^{d}$} Edward Jager,\textit{$^{d}$}  Emma Springate,\textit{$^{e}$} Oliver Alexander,\textit{$^{e}$} Cephise Cacho,\textit{$^{e}$} Richard T. Chapman,\textit{$^{e}$}, Adam Kirrander,\textit{$^{a}$} and Russell S. Minns\textit{$^{d}$}}

\address{\textit{$^{a}$EaStCHEM, School of Chemistry, University of Edinburgh, David Brewster Road, Edinburgh EH9 3FJ, United Kingdom.}\\
\textit{$^{b}$~Center for Free-Electron Laser Science, DESY, Notkestrasse 85, 22607 Hamburg, Germany}\\
\textit{$^{c}$The Hamburg Centre for Ultrafast Imaging, University of Hamburg, Luruper Chaussee 149, 22761 Hamburg, Germany}\\
\textit{$^{d}$~Chemistry, University of Southampton, Highfield, Southampton SO17 1BJ, UK}\\
\textit{$^{e}$Central Laser Facility, STFC Rutherford Appleton Laboratory, Didcot, Oxfordshire OX11 0QX, UK }}

\begin{abstract}
%% Text of abstract (Max 100 words)
{New \it{ab-initio}} surface hopping simulations of the excited state dynamics of CS$_2$ including spin-orbit coupling are compared to new experimental measurements using a multiphoton ionisation probe in a photoelectron spectroscopy experiment. The calculations highlight the importance of the triplet states even in the very early time dynamics of the dissociation process and allow us to unravel the signatures in the experimental spectrum, linking the observed changes to both electronic and nuclear degrees of freedom within the molecule.
\end{abstract}
\maketitle

\section{Introduction}
\label{S:1}
The dissociation dynamics of CS$_2$ following UV excitation have been a benchmark in chemical dynamics for many years, with numerous experimental studies in both the time and frequency domain, see for example \cite{Townsend06,Kitsopoulos01,Brouard12,spesyvtsev2015,Bisgaard09,Horio13,Horio14,Hemley83,Farmanara:JCP111:5338,Hockett:NP7:612}. This lasting fascination with CS$_2$ can be traced to the efficient dissociation, dictated by complex dynamics on multiple coupled electronic states. Despite intense experimental study, the fast dynamics and the high ionisation limits of intermediates and final products have limited the experimental view to specific points along the full dissociation path, such that open questions remain even for this structurally simple molecule.  The origin of the complexity derives from the near degeneracy of the optically bright $^1$B$_2$($^1$$\Sigma_u^+$) state with multiple other electronic states at linear geometry, which leads to highly efficient population transfer and strongly coupled multistate dynamics. The mixing of the electronic states leads to dissociation and the formation of a ground state CS (X $^1\Sigma^+$) molecule in conjunction with atomic sulphur in either the spin forbidden ground state, $^3$P, or a spin allowed excited state, $^1$D. While the exact branching ratio has proven difficult to define accurately, the spin forbidden product is seen to dominate in most experimental studies \cite{Waller87,Kitsopoulos01,Xu04}, highlighting the importance of spin-orbit coupling for an accurate description. Considering how well studied this molecule has been experimentally,  calculations of the dynamics have been limited with, as far as we are aware, no simulations accounting for the spin-orbit coupling that drives the dominant dissociation process. In this work we combine {\it{ab-initio}} surface hopping simulations of the dissociation dynamics of CS$_2$ with new photoelectron spectroscopy measurements using a multiphoton probe to study the effect of spin-orbit coupling on the early time dynamics of the molecule.

Previous dynamics calculations have focused on the singlet state dynamics and the effect of non-adiabatic coupling on measured photoelectron angular distributions obtained following excitation at 201 nm and ionisation with 268 nm \cite{Schuurman2014}. These calculations provide a very good measure of the photoangular distribution, which they claim suggests that the lack of spin-orbit coupling in the model does not affect the calculated early time dynamics. It should be noted that the rather low energy probe used in the experiment means that only the singlet states could be ionised and that any effect of the triplet in the angular distributions would not be observable. The measurements and theory therefore do not take into account the population transfer between the initially excited singlet manifold and the accessible triplet states. This point is highlighted by recent time-resolved VUV photoelectron spectroscopy experiments by Spesyvtsev {\it{et al.}}\ \cite{spesyvtsev2015} using a 20 fs 159 nm (7.8 eV) probe pulse. These experiments provide the most detailed maps of the excited state dynamics to date, and show large changes in electron kinetic energy as the molecule undergoes bending vibrations with an almost 3 eV shift in the measured electron kinetic energy in 40 fs. The probed dynamics occur on the singlet surfaces and, as the molecule continues to vibrate, the population is transferred  into lower lying electronic states which are outside the observation window provided even by their VUV probe.

\section{Theory}
\label{S:2}

\subsection{Computational methods}

\begin{figure}[ht!]
\centering
\includegraphics[width=8.6cm]{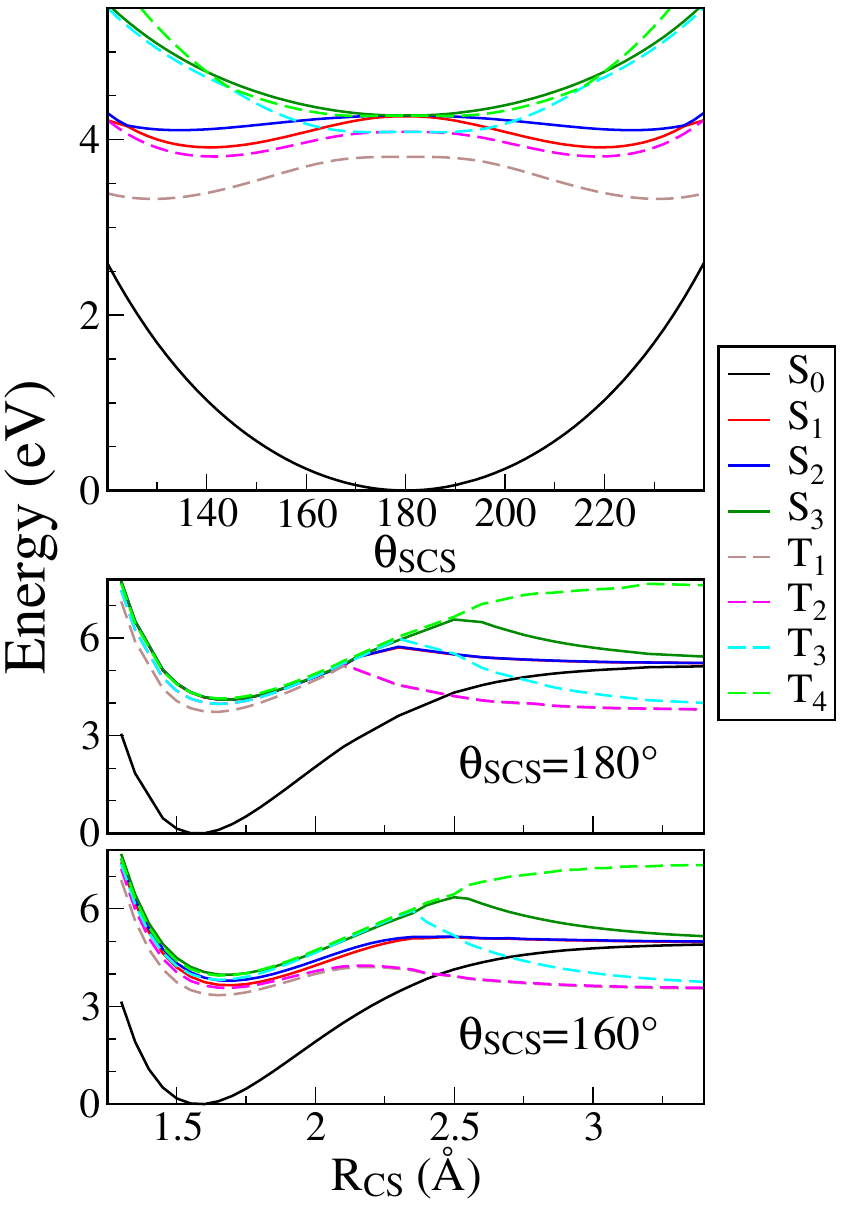}
\caption{Bending $\theta_{\text{SCS}}$ (top), and radial asymmetric stretch $R_{\text{CS}}$ for linear $\theta_{\text{SCS}}$=\ang{180} (middle)
  and bent $\theta_{\text{SCS}}$=\ang{160} (bottom), potential energy curves calculated at the
  SA8-CAS(16,12)/aug-cc-pvTZ level for the first four singlet and
  triplet states of CS$_2$. Remaining degrees of freedom are frozen at
  their equilibrium values.}
\label{fig:potentials}
\end{figure}

{\it{Ab-initio}} electronic structure calculations were performed
using the Molpro 2015.1 \cite{MOLPRO_brief} suite of programs at the
CAS(16,12)/aug-cc-pvTZ and aug-cc-pvQZ level of theory using the full
valence (16,12) active space with bonding and antibonding $\sigma$ and
$\pi$ molecular orbitals and sulfur atom lone pairs. Ground state
geometry optimisation using CAS(16,12)/aug-cc-pvQZ resulted in
$R_{\text{CS}}$=\SI{1.569}{\angstrom} and
$\theta_{\text{SCS}}$=\ang{0}. Angular and
radial cuts through the singlet and triplet potential energy surfaces are
shown in Fig.\ \ref{fig:potentials}, and vertical excitation energies and
oscillator strengths for the first four excited singlet
states are given in Table \ref{tab:excitation}. The potential energy curves in Fig.\ \ref{fig:potentials} are broadly in keeping with previous {\em{ab-initio}} calculations \cite{tseng1994abinitio,zhang1995ab,brown1999excited,wiberg2005comparison,
mank1996detailed}.

\begin{table}[ht!]
\centering
\caption{Vertical excitation energies ($\triangle E=E(\text{S}_i)-E(\text{S}_0)$) and oscillator strengths from the ground state to the first four excited singlet states of CS$_2$,
  calculated using SA5-CAS(16,12)-SCF/aug-cc-pvQZ with CASPT2
  corrections to the energies.
  The excitation energies are calculated at the equilibrium geometry
  ($\theta_{\text{SCS}}$=\ang{180} and $R_{\text{CS}}$=\SI{1.569}{\angstrom}), while
  oscillator strengths are calculated at
  $\theta_{\text{SCS}}$=\ang{160} since transition
   are very weak in the linear geometry.
}
\begin{tabular}{|l|c|c|}
\hline
State  & Energy (eV) & Oscillator strength   \\ \hline
S$_1$ & $3.821$   & $0$ \\
S$_2$ & $3.836$   & $0.004282$ \\
S$_3$ & $3.836$   & $0$ \\
S$_4$ & $6.430$   & $0.000834$ \\
\hline
\end{tabular}
\label{tab:excitation}
\end{table}

We simulate the dynamics of photoexcited CS$_2$ using the code SHARC
\cite{sharc-md,Richter2011JCTC} interfaced with MOLPRO
\cite{MOLPRO_brief}.  SHARC treats nuclear motion classically, but
nonadiabatic effects and spin-orbit coupling \cite{Mai2015IJQC} are
included using the fewest-switches surface-hopping approach
\cite{tully1990molecular}. The spin-orbit coupling is treated in the diabatic representation and has been shown to replicate branching dynamics in IBr when compared to full quantum dynamics simulations \cite{Mai2015IJQC}. In contrast to previous singlet-only simulations \cite{Schuurman2014}, we propagate the dynamics on the four
lowest singlet {\em{and}} triplet electronic states. To keep the
simulations computationally feasible, we perform the electronic structure calculations
at the SA8-CAS(8,6)-SCF/6-31G* level, which qualitatively reproduces
the potential energy curves shown in Fig.\
\ref{fig:potentials}. The differences to the CAS(16,12)/aug-cc-pvTZ level calculations are minor at small and large bond-lengths, but at intermediate distances the smaller active space gives rise to elevated barriers to dissociation, that lead to transient trapping of population in the T2 state (see discussion in Section 2.2). Initial positions are generated from a Wigner
distribution based on the CAS(8,6)/6-31G* ground state vibrational
frequencies and the oscillator strength of each geometry, and kinetic
energy is assigned based on the required excitation energy and the
experimental pump pulse energy.  Following
this protocol, 85\% of trajectories begin in the
\textit{B} $^1$B$_2$ state. A total of
369 trajectories are launched, of which 197 reach 500 fs and 114 reach 1000 fs, using a time step
of 0.5 fs. The reasons that some trajectories fail to reach 1000 fs are related to the electronic structure calculations and include numerical problems such as excessive gradients in the CI or failure in convergence of the MCSCF calculations.

\subsection{Computational results}

\begin{figure}[!ht]
  \subfloat[Bondlengths $R_{\text{CS}}$ as a function of time.\label{fig:bondlengths}]{
       \includegraphics[width=8.6cm]{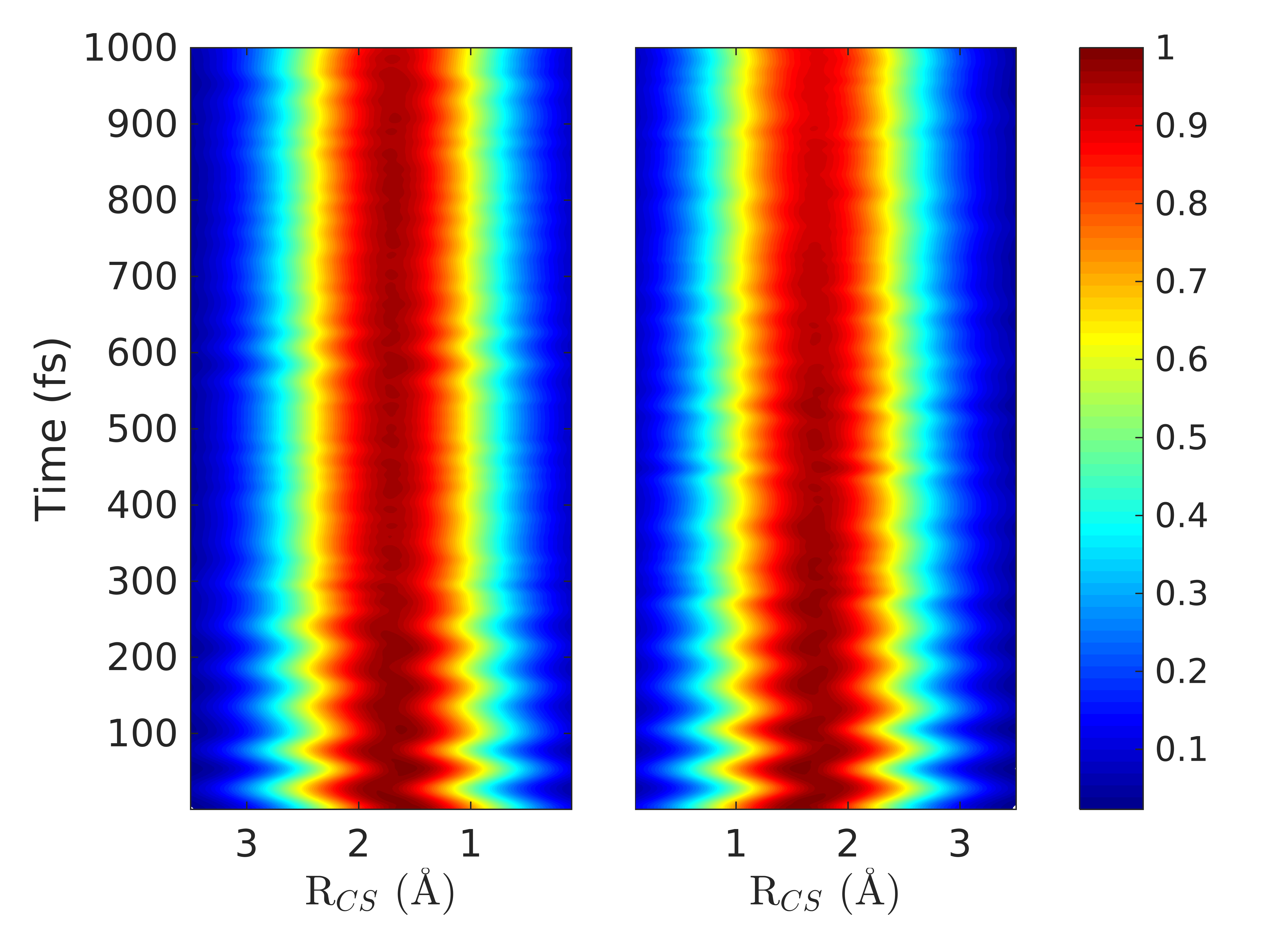}%D1_D2_1000fs.png
     }
%     \hfill
     \subfloat[Bending angle $\theta_{\text{SCS}}$ as a function of time.\label{fig:angle}]{
       \includegraphics[width=8.6cm]{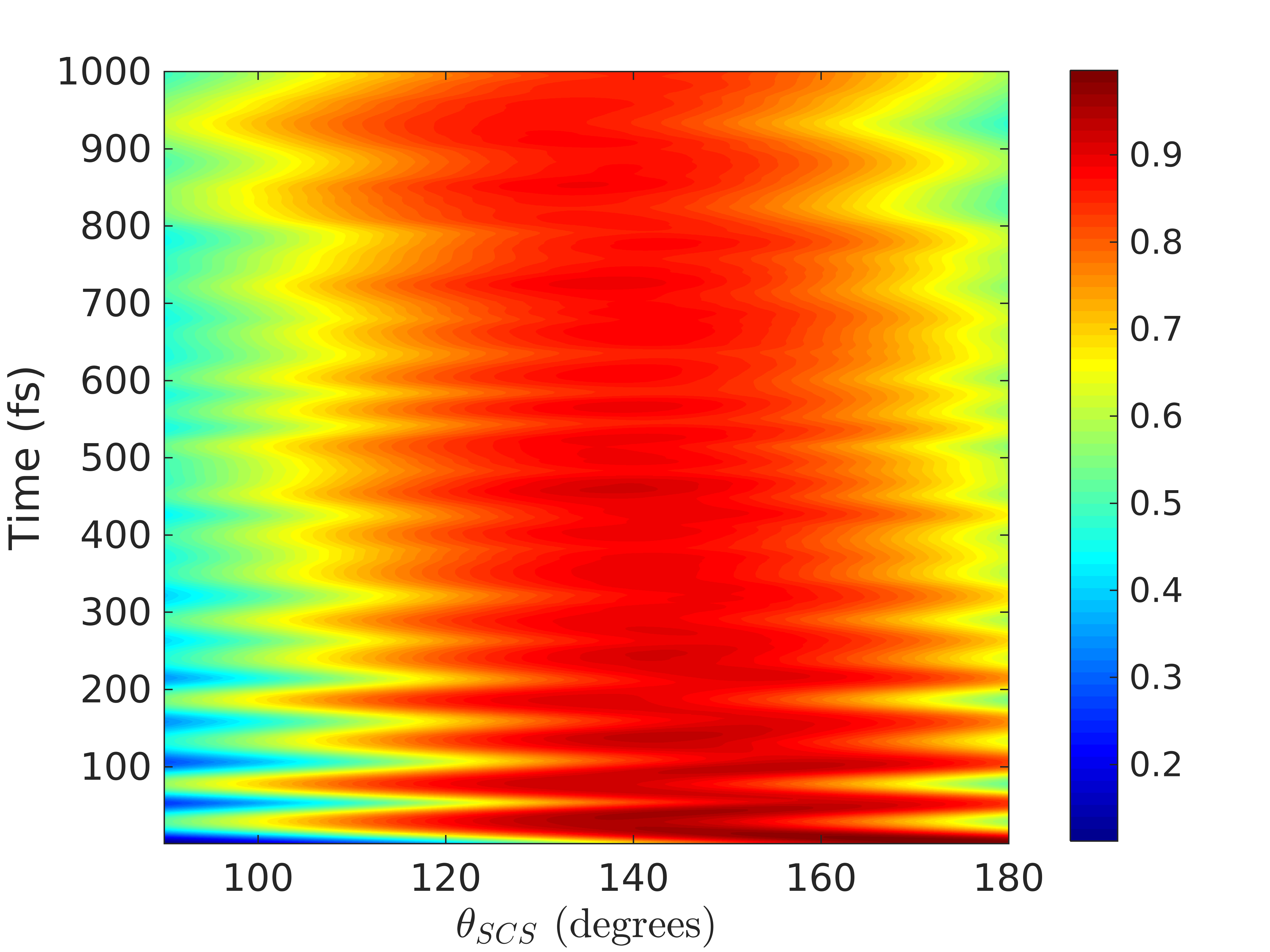} % angle_1000fs.png
     }
     \caption{Probability density evolution of the CS$_2$ geometry in
       terms of the two bondlengths $R_{\text{CS}}$ (Fig.\
       \ref{fig:bondlengths}) and bending angle $\theta_{\text{SCS}}$
       (Fig.\ \ref{fig:angle}) from the simulations.}
     \label{fig:dynamics-theory}
   \end{figure}

\begin{figure}[ht!]
\centering
\includegraphics[width=8.6cm]{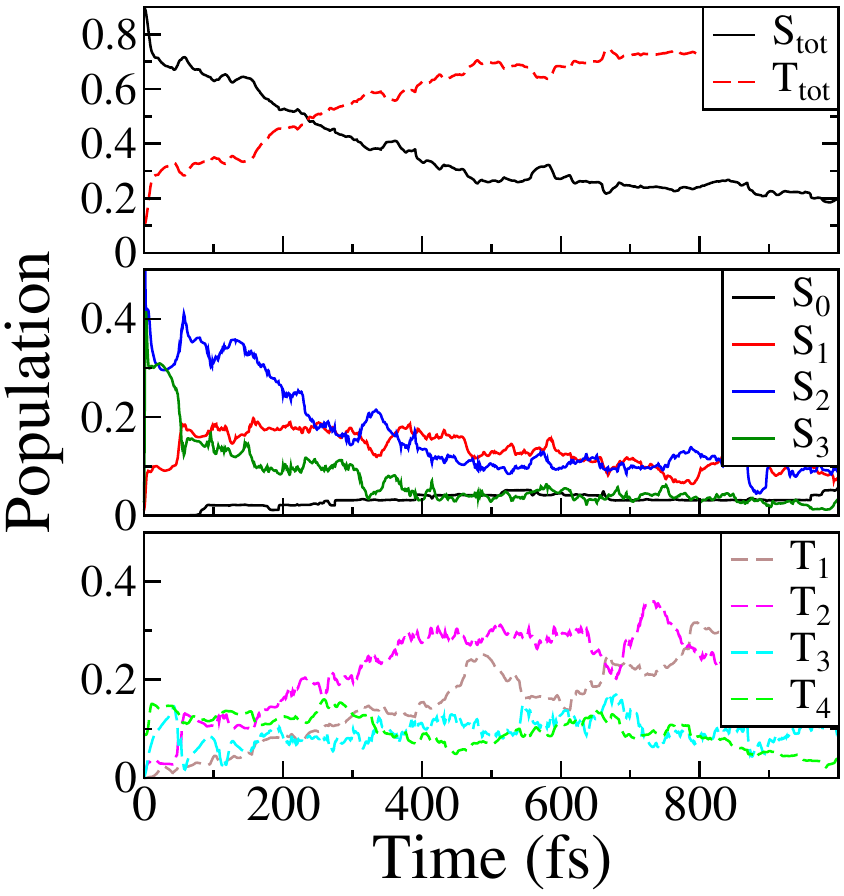}
\caption{Time-dependent adiabatic state populations from simulation of CS$_2$ dynamics. The top panel shows the total singlet (excluding the S$_0$ ground state) and total triplet populations as a function of time. The middle panel shows the populations for each singlet state and the bottom panel the corresponding
data for the triplet states.}
\label{fig:population}
\end{figure}

Excitation of CS$_2$ triggers bending and vibrational motion in the molecule, as can
be seen in Fig.\ \ref{fig:dynamics-theory}, which shows the probability
density evolution of the molecular geometry as a function of the C-S
bond-lengths, $R_{\text{CS}}$, and the bending angle,
$\theta_{\text{SCS}}$.  During the first $100$ fs the vibrations are
dominated by the symmetric stretch, but at later times energy flows
into the asymmetric stretch. The frequencies of vibrations are
somewhat over-estimated compared to the experimental values,
presumably due to slight differences in the {\it{ab-initio}} potential energy surfaces at
the CAS(8,6) level. The total fraction of dissociated molecules in the full set of 369 trajectories is 22\%, which constitutes a lower bound since only about a quarter of the trajectories reach 1000 fs. Dissociation occurs predominantly in the triplet states, with  89\% of the trajectories that dissociate occurring on the triplet surfaces. The lower
degree of dissociation compared to the experiments can be traced to the topology of the potential energy surfaces at the
level of {\it{ab-initio}} theory employed in the simulations, as
discussed below.

The electronic state populations as a function of time are shown in
Fig.\ \ref{fig:population}.  Initial excitation onto the S$_2$
$^1\text{B}_2$($^1\Sigma_u^+$) state is followed by rapid decay onto
the singlet S$_3$ and S$_1$ potentials, as well as a redistribution of
population onto the manifold of triplet states via spin-orbit
coupling. The nonadiabatic transfer of population between the singlet
states correlates strongly with the bending motion of the molecule,
with efficient transfer predominantly occurring close to the linear
geometry where states are (near)-degenerate. This gives rise to a periodic beating in both the individual singlet state populations and in the total singlet population. Over time there is
a build-up of population in T$_2$ at $t>400$ fs, and a subsequent rise
of population in T$_1$ at around $t>800$ fs, due to population transfer
from T$_2$ to T$_1$. The build-up in T$_1$ appears to be an artifact due to the SA8-CAS(8,6)-SCF/6-31G* {\it{ab-initio}} calculations, which increases the relative barrier height for dissociation on the T$_1$ and T$_2$ potentials by $\approx1.5$ eV (see Section 2.1), hindering dissociation and
leading to the observed accumulation of population in
T$_2$. Consequently, it is reasonable to assume that the population
trapped in T$_2$ in actual fact dissociates as
observed in the experiment. Nevertheless, despite that the simulations
underestimate the amount of $t<1$ ps dissociation via the triplet
states, the short-time $t<400$ fs dynamics appears quite reliable.

\section{Experiment}
\subsection{Experimental methods}
The experiment has been described in detail previously \cite{smith2016}. Briefly, an amplified femtosecond laser system (Red Dragon, KM Labs) generates 30~fs pulses of 800~nm light, with a pulse energy of up to 10~mJ at a repetition rate of 1~kHz. The pump pulse is produced via fourth harmonic generation of the fundamental (800~nm) beam, generating photons at around 200~nm. The 200~nm beam is produced using standard non-linear optics with sequential second, third and fourth harmonic generation in BBO giving a pulse energy of $\sim$1~$\mu$J. The 400~nm probe is generated by second harmonic generation of the fundamental laser output, producing approximately 5~$\mu$J per pulse. The pump and probe beams are reflection focused in a near collinear geometry and cross at the centre of the interaction region of a velocity-map imaging (VMI) spectrometer~\cite{Eppink1997}, where they intersect the CS$_2$ molecular beam. The pump and probe beams are both linearly polarised in the plane of the VMI detector, perpendicular to the time-of-flight axis. The molecular beam is generated through the expansion of 5\% CS$_2$ in He at  1~bar through a 1~kHz pulsed nozzle (Amsterdam cantilever \cite{Janssen2009}) with a 100~$\mu$m aperture. The resulting expansion passes through a 1~mm skimmer and enters the interaction region of the spectrometer through a hole in the centre of the repeller plate of the VMI spectrometer. The photoelectron spectra are obtained through polar onion-peeling of the background subtracted images~\cite{POP}. Although the photoelectron angular distributions are obtained, they show no time dependence and as such are not discussed in the results section.

\subsection{Experimental results}

\begin{figure}
	\centering
	\includegraphics[width=0.5\linewidth]{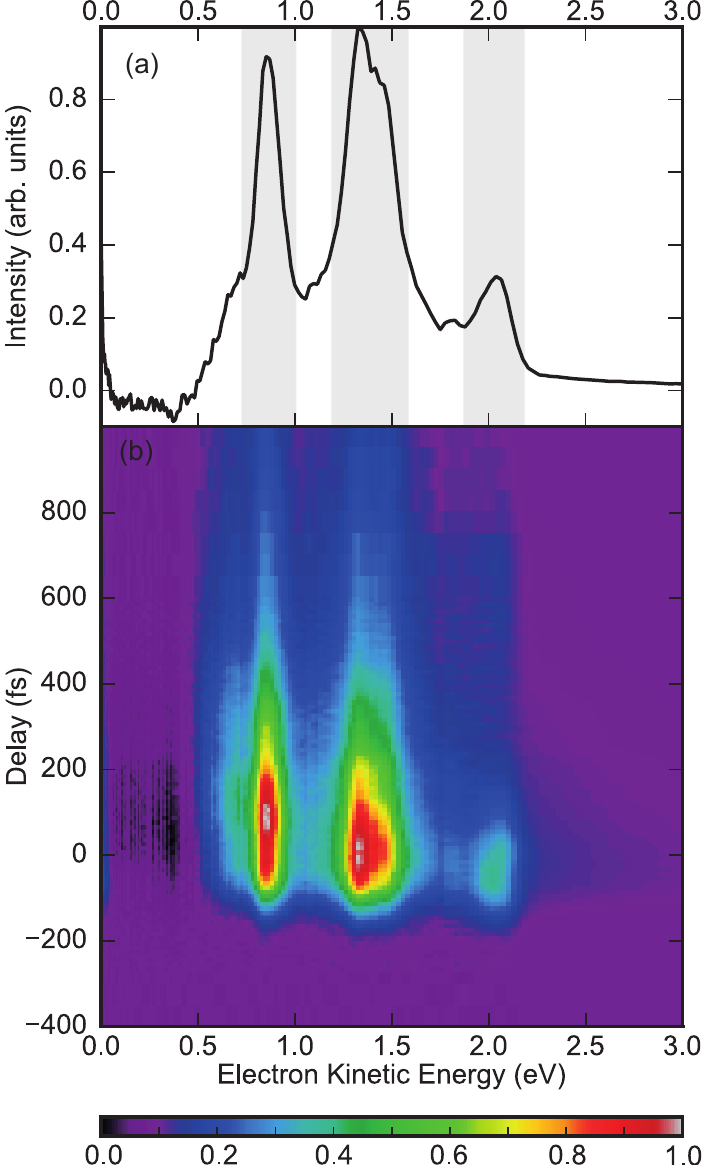}
	\caption{(a) Photoelectron spectrum obtained when the pump and probe pulses are temporally overlapped. The highlighted regions mark those used in the integrated plots shown in figure \ref{fig:dynamics}. (b) Photoelectron spectra as a function of pump-probe delay.}
	\label{fig:combined_spectrum}
\end{figure}

The 200~nm pump excites a vibrational wavepacket, predominantly in the S$_2$ $^1$B$_2$ excited electronic state. The motion is then probed by non-resonant two-photon absorption at 400~nm. This provides a total energy of 12.5~eV, with the ionisation potential of CS$_2$ at 10.07~eV. The photoelectron signal obtained when the pump and probe pulse are overlapped in time is plotted in Fig.\ \ref{fig:combined_spectrum}(a) with three main features around 2.1~eV, 1.4~eV and 0.9~eV electron kinetic energy. The spacing between the features is similar to that seen in previous single-photon ionisation measurements~\cite{Townsend06,Bisgaard09}. The use of a multiphoton probe maintains a clean experimental measurement, without any probe-pump contributions at early times, while maximising the available energy for ionisation, such that we can observe much of the initial excited state dynamics.
\begin{figure}
	\centering
	\includegraphics[width=0.4\linewidth]{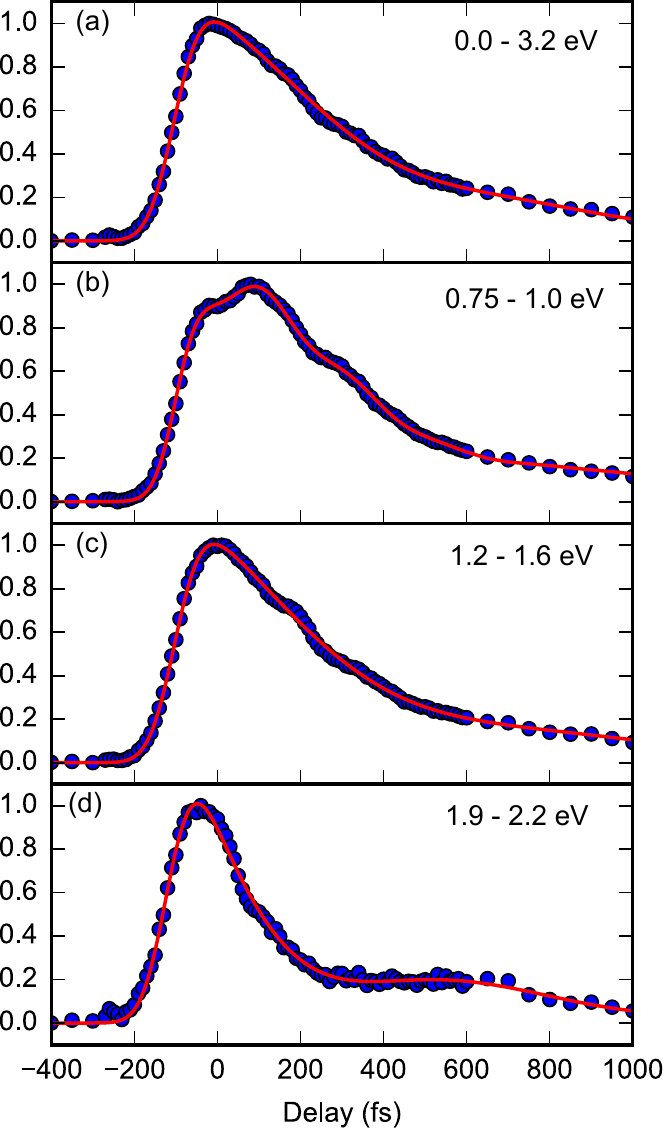}
	\caption{Total integrated photoelectron intensity (a) and intensity within the shaded areas shown in Fig. \ref{fig:combined_spectrum}; 0.75 - 1.00~eV (b), 1.20 - 1.60~eV (C) and 1.90 - 2.20~eV (d). Solid lines represent fits to the data. (a,c,d) were fit with a single exponential decay, modulated by a single damped oscillator and convoluted with the instrument response function. However, a second oscillation with a period $\sim$200~fs is clearly visible in the data. The data in (b) was fit with two damped oscillating components, which reproduce both observed oscillatory features well.}
	\label{fig:dynamics}
\end{figure}

The time-dependence of the photoelectron spectrum is shown in Fig.\ \ref{fig:combined_spectrum}(b). The three peaks in the spectrum have different appearance times, with those at lower electron kinetic energy appearing after those at higher electron kinetic energy. The low-energy feature rises approximately 35~fs after the highest energy feature at 2.1~eV. This maps the initial bending motion as seen in the calculated dynamics, Fig.\ \ref{fig:angle}, and in previous measurements~\cite{spesyvtsev2015}. At longer delay times it is also clear that the centre of mass of the photoelectron spectrum shifts to lower electron kinetic energies, such that the lifetime of the measured photoelectron features is longer at lower electron kinetic energies. To obtain a clearer view of the changes observed at the various electron kinetic energies measured, we plot the integrated intensity over the features highlighted in Fig.\ \ref{fig:combined_spectrum}(a) in Fig.\ \ref{fig:dynamics}. The difference in lifetime is apparent in the plots, as well as the appearance of clear oscillations in intensity that peak at times after time zero. The effect of the oscillations is most prominent in the feature centered around 0.9~eV, Fig.\ \ref{fig:dynamics}(b), which has a maximum intensity $\sim$200~fs after excitation. None of the transients can therefore be fit to a simple exponential decay, but are modulated by at least one oscillating component. We therefore fit the transients to an exponential decay modulated by either one or two damped oscillations, convoluted with the instrument response function, corresponding to the laser pulse cross-correlation~\cite{Hanggi:AnalChem57:2394},

\begin{equation}
 \label{fits}
g\otimes\left(A_0\exp(-\frac{t-t_0}{\tau})\times \prod^n A_n\cos(\omega_n(t-t_0)+\delta_n)\right) .
 \end{equation}

Here  $A_n$ represent intensity scaling parameters, $t_0$ the arrival time of the laser pulse, $\tau$ the exponential lifetime and $\omega$ and $\delta$ the angular frequency and phase of the oscillatory component. Fits are plotted as solid lines in Fig.\ \ref{fig:dynamics}. The highest energy feature, Fig.\ \ref{fig:dynamics}(d), provides the clearest data set and contains a single oscillation of period $\sim$0.9 ps, 38~cm$^{-1}$, as has previously been experimentally observed~\cite{Farmanara:JCP111:5338,Hockett:NP7:612}. This corresponds to the beat between the $\nu_1$ and $\nu_2$ vibrational modes~\cite{Hemley83}. This is present in each of the other features in the spectrum, along with a second beat with a period around 200~fs. The effect of this oscillation is clearest in the trace presented in Fig.\ \ref{fig:dynamics}(b), however the mixing with the other oscillation and relatively low contrast makes assigning the absolute value of this oscillation difficult, leading to significant error margins. Nonetheless we extract an oscillation period of 220~fs, corresponding to 149~cm$^{-1}$, from this data. While this oscillation period does not fit with any of the known vibrational periods of the molecule, similar frequencies were also observed in a previous study~\cite{spesyvtsev2015} but were not discussed or assigned. The fits to the experimental data furthermore yield an increase in lifetime towards the lower electron kinetic energy regions. The $1/e$ lifetimes extracted are 401~fs, 452~fs and 457~fs for the peaks at 2.1~eV, 1.4~eV and 0.9~eV respectively.

\section{Discussion}
\begin{figure}[ht!]
  \subfloat[Bond-lengths $R_{\text{CS}}$.\label{fig:bondlengths-singlet}]{
       \includegraphics[width=7.1cm]{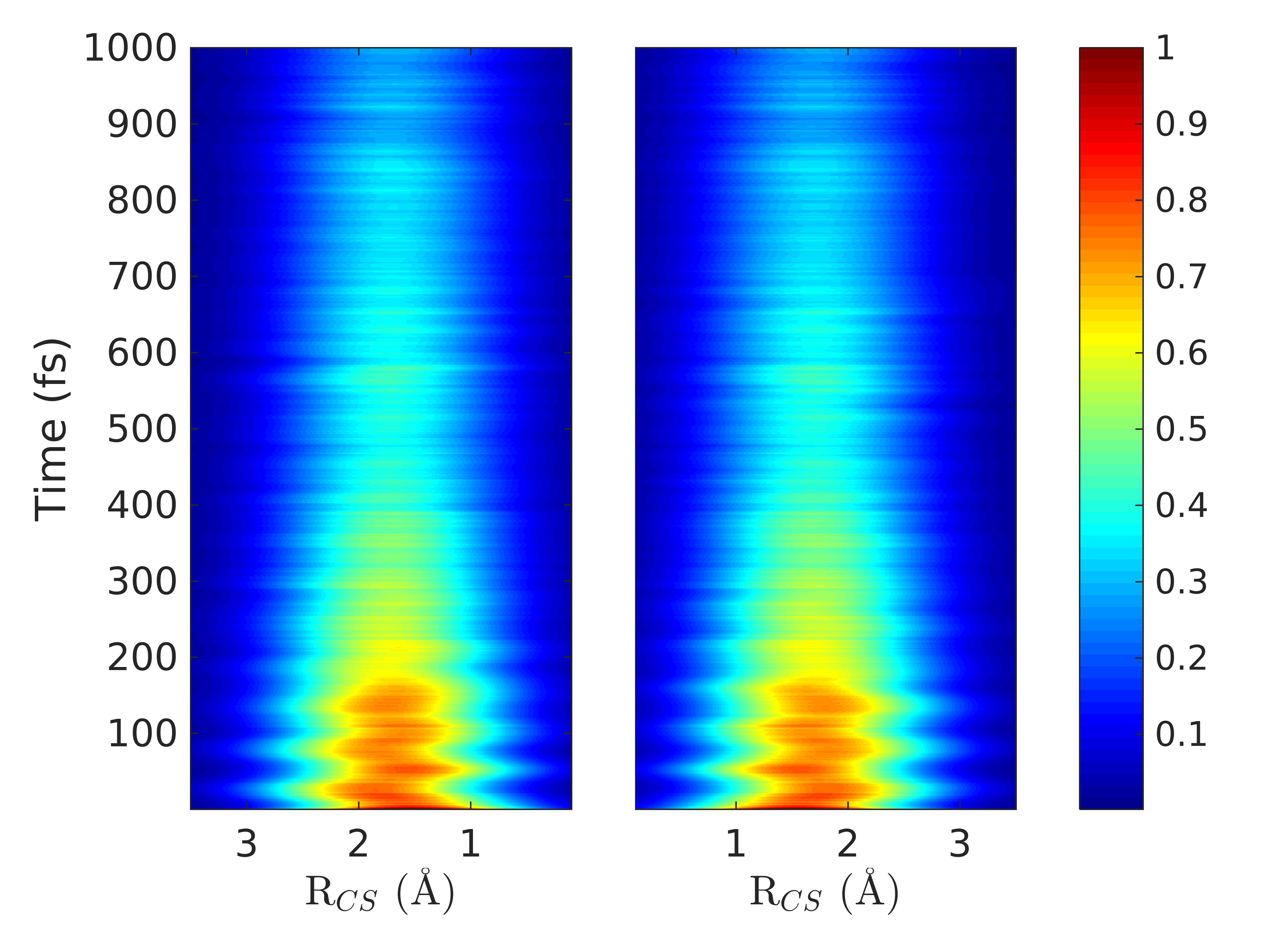}
     }
     \subfloat[Bending angle $\theta_{\text{SCS}}$.\label{fig:angle-singlet}]{
       \includegraphics[width=7.1cm]{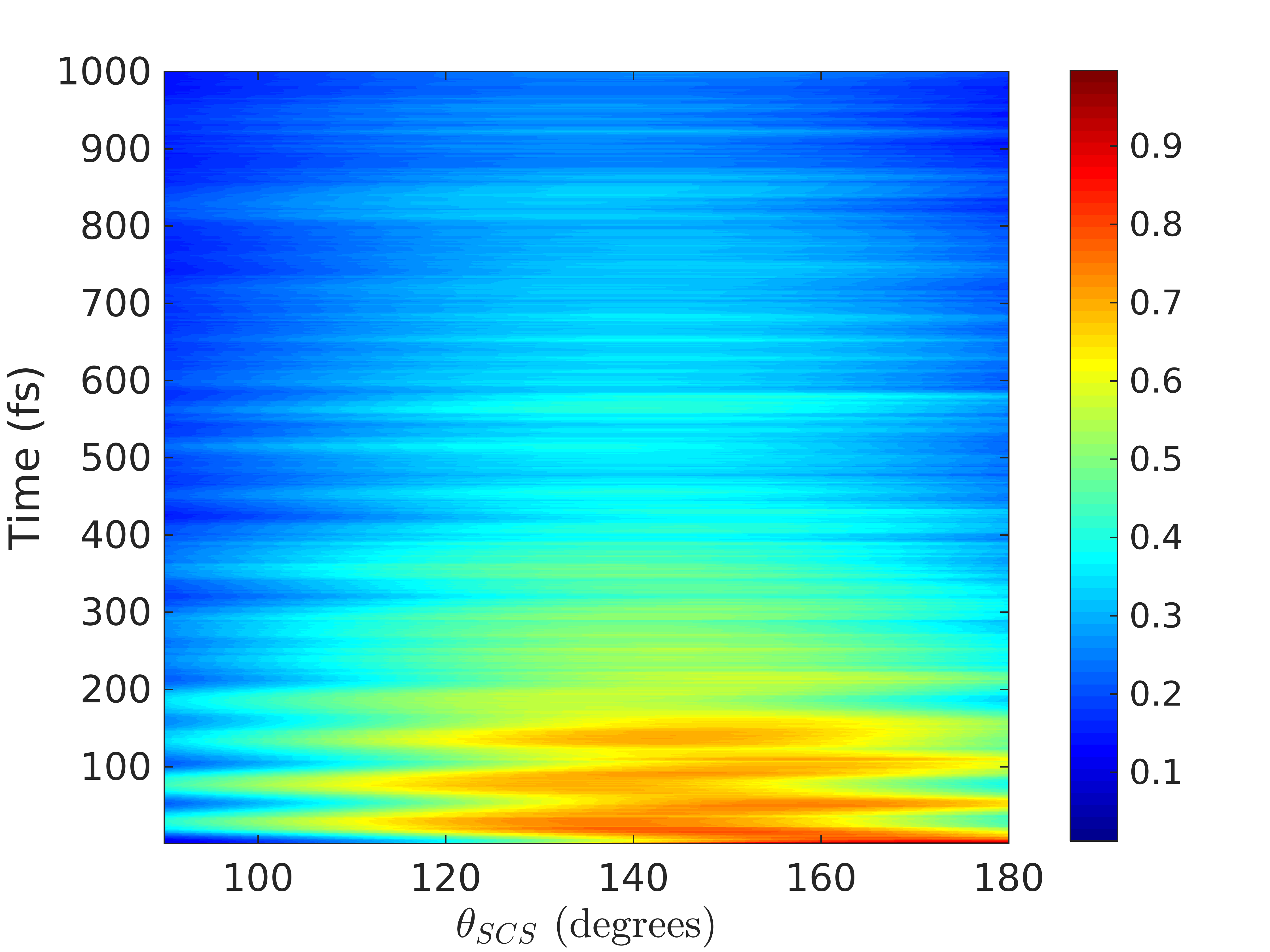}
     }
     \caption{Probability density evolution of the CS$_2$ geometry in terms of the two bond-lengths $R_{\text{CS}}$ and bending angle $\theta_{\text{SCS}}$ in the singlet states only. The intensity bar shows the total population with the decreasing intensity showing the transfer of population into the triplet states.}
     \label{fig:combined_singlet}
   \end{figure}

We now provide a comparison of the experimental measurements and the theoretical calculations. For both the calculations and experiment it is clear that the triplet states play a large role in the dynamics from very early times. Significant population is transferred into the triplet states very rapidly with over 50\% of the total population in the triplet states within 250~fs. The overall transfer of population approximately matches the decay rate measured in the experiment such that we are only sensitive to the singlet state population. As mentioned above the initial shift in the measured electron kinetic energy maps the initial bending motion of the molecule. As the pump-probe delay increases, the electron kinetic energy shifts towards lower values, such that we observe a longer lifetime for the lower electron kinetic energy regions in the spectrum in Fig.\ \ref{fig:dynamics}. To compare the measured signal to theory, in Fig.\ \ref{fig:combined_singlet} we plot the singlet state component of Fig.\ \ref{fig:dynamics-theory}. Initial excitation leads to a wavepacket that oscillates between linear geometries and an angle of $\sim$110$^\circ$. With increasing pump-probe delay, the range of angles explored narrows and moves away from the linear geometries associated with the spectral feature at the highest electron kinetic energy. Within the current experiment we do not have the time resolution to fully resolve the bending motion, but we do observe the effect of the narrowing and shifting of the angles explored by the molecule as a corresponding narrowing and shifting to lower electron kinetic energies in the photoelectron spectrum.

As mentioned above, the short time period oscillation seen both in the total photoelectron count rate and, in individual regions of the photoelectron spectrum (Fig. \ref{fig:dynamics}), do not match any vibrational periods within the molecule. The simulations also show no obvious changes in the molecular structure that appear to provide an explanation for the oscillations. The calculations do show periodic changes in total singlet excited state population that correlate with the observed changes the photoelectron yield. We therefore tentatively assign the oscillations in the experimental spectrum to changes in the total singlet state population.

\section{Summary}
We have performed a combined theory and experiment study of the excited state dynamics of CS$_2$. The ab-initio surface-hopping simulations highlight the importance of the triplet states in the early time dynamics with significant population transfer predicted, and observed in the complementary time-resolved photoelectron spectroscopy measurements. The combined work demonstrates that one can now do on-the-fly dynamics including spin-orbit coupling. The accuracy of the calculation is such that we are able to directly compare the results of the calculation with experiment and explain the shifting and narrowing of the photoelectron spectrum in terms of the bending motion and angles explored by the vibrational wavepacket, while oscillation in the measured photoelectron count rate are explained by the complex coupling of the electronic states that leads to rapid population transfer between manifolds of multiple singlet and triplet excited states. Measuring the longer term dynamics in the triplet states is an experimental challenge. Recent experiments by the Suzuki group using a 7.8 eV probe showed no clear contributions from the triplet states suggesting ionisation requires higher energies.\cite{spesyvtsev2015} This is presumably due to the ionisation propensity of the triplet states being into electronically excited ion states. Measurements of the dynamics outside of the initially excited singlet states will therefore require measurement using a significantly higher photon energy such as that available from a high harmonic generation source. Such experiments are currently ongoing and will be the subject of a future publication.

\section*{Acknowledgments}
All authors thank the STFC for access to the Artemis facility (app. number 13220015). RSM thanks the Royal Society for a University Research Fellowship (UF100047) and the Leverhulme trust for research support and for ADS's studentship (RPG-2013-365). HMW thanks the Central Laser Facility and Chemistry at the University of Southampton for a studentship. EJ thanks Chemistry at the University of Southampton for a studentship. We also acknowledge funding from the EC's Seventh Framework Programme (LASERLAB-EUROPE, grant agreement n$^\circ$ 228334). We thank Phil Rice for technical assistance. This work has been supported by the excellence cluster "The Hamburg Center for Ultrafast Imaging -- Structure, Dynamics and Control of Matter at the Atomic Scale" of the Deutsche Forschungsgemeinschaft (CUI, DFG-EXC1074). D.A.H. was supported by the European Research Council through the Consolidator Grant K{\"{u}}pper-614507-COMOTION. AK acknowledges funding from the
European Union (FP7-PEOPLE-2013-CIG-NEWLIGHT) and the Leverhulme Trust (RPG-2013-365), and DB acknowledges a PhD studentship from the University of Edinburgh. The computational work reported used the ARCHER UK National Supercomputing Service
(http://www.archer.ac.uk) and the Edinburgh Compute and Data Facility (ECDF) (http://www.ecdf.ed.ac.uk). DB thanks Sebastian Mai (Wien) for helpful discussions.

\bibliographystyle{model1-num-names}
\bibliography{CS2.bib}
\end{document}